\documentclass[prp,aps,twocolumn,preprintnumbers,floats,amsmath,amssymb]{revtex4}
\usepackage{graphicx}
\usepackage{dcolumn}
\usepackage{bm}
\usepackage{amssymb}
\usepackage[latin1]{inputenc}

\newcommand{\GA}{G^{[\cal A]}}
\newcommand{\GAm}{G^{[{\cal A}-1]}}

\def \be{\begin{equation}}
\def \ee{\end{equation}}

\begin{document}
\title{A knitting algorithm for calculating Green functions in quantum systems}

\author{K.Kazymyrenko}
\author{X.Waintal}
\affiliation{Service de Physique de l'Etat condens\'e,\\
DSM/DRECAM/SPEC, CEA Saclay\\
91191 Gif sur Yvette Cedex, France}
\begin{abstract}
We propose a fast and versatile algorithm to calculate local and transport properties such as conductance, shot noise, local density of state or local currents in  mesoscopic quantum systems. Within the
non equilibrium Green function formalism, we generalize the recursive Green function technique to tackle
multiterminal devices with arbitrary geometries. We apply our method to analyze two recent experiments: an electronic 
Mach-Zehnder interferometer in a 2D gas and a Hall bar made of graphene nanoribbons in quantum Hall regime. In the latter case, we find that the Landau edge state pinned to the Dirac point gets diluted upon increasing carrier density.
\end{abstract}

\maketitle


The field of quantum transport at the nanometer scale now includes a large number of systems involving very different physics. Examples include for instance mesoscopic devices in two dimensional heterostructures~\cite{metal05b}, graphene nanoribbons~\cite{rycer07}, superconducting weak links~\cite{gunse94}, molecular electronic devices~\cite{damle01} or ferromagnetic multilayers nanopillars~\cite{sanvi99,haney07}. Although 
those systems have different structures and geometries, they are all quantum systems connected to the macroscopic world through electrodes and, consequently, formalisms developed to describe one of them can often be adapted to the others. This is in particular the case of the widely used Landauer-B\"uttiker formalism~\cite{blant00} which focuses on the scattering properties of the system. The formalism is very intuitive and general. However, it is not well suited when one is interested in what happens {\it inside} the sample or for performing a microscopic calculation for a given device. An alternative mathematically equivalent approach is referred as the NEGF (non equilibrium Green function) formalism ~\cite{carol71,meir92}. NEGF, which is derived from the Keldysh formalism~\cite{keldy64}, provides a simple route to compute the physical observables from a microscopic model. 
It is now an extremely popular numerical approach to a very wide class of physical problems (see references in~\cite{lake06}). For instance, all the references mentioned in examples above correspond to calculation done with  this technique either from ab-initio or from phenomenological models~\cite{metal05b,rycer07,gunse94,damle01,brand02,sanvi99,khomy05,haney07}. At the core of NEGF is the calculation of the retarded Green function $G$ of the mesoscopic region in presence of the (semi-infinite) electrodes. 
A straightforward method consists of direct inversion of the Hamiltonian $H$. However, when doing so, one is restricted to rather small systems of a few thousand sites:
for a system of size $L$ in dimension $d$, the computing time scales as $L^{3d}$ while the
needed memory scales as $L^{2d}$.  An alternative algorithm, known as the recursive Green function 
technique~\cite{macki85,sanvi99,metal05b}, takes advantage of the structure of $H$ to reduce drastically the computing time down to $L^{3d-2}$, putting systems of a few million sites within reach. In its original version~\cite{macki85,lee81,thoul81},
only the transport properties of the device could be computed, but recent progresses made it possible to get access to observables inside the sample (like local electronic density or local currents~\cite{metal05b,crest03}) at a cost  $L^{2d-1}$ in memory. The recursive Green function technique suffers however from a serious limitation: in its original formulation it is intrinsically one dimensional and most applications are done for quasi-one dimensional bars connected to two electrodes. On the other hand, real devices often have more than two electrodes and more complicated geometries. 
This paper is devoted to a general versatile algorithm to simulate multiterminal systems
with arbitrary geometries and topologies.
Earlier works in this direction are scarce. Baranger and al.~\cite{baran91} considered the Hall effect in a 2D cross (a specific code was developed to handle this geometry). Modular algorithms~\cite{sols89,rotte03} allow
to compute the properties of 2D quantum ballistic billiards. Most references with multiterminals involve direct inversions with small system sizes; others adaptation of the algorithm to the specific problem at 
hand~\cite{guan03,polin06}. Competitors to the present algorithm are also being developed~\cite{wimme07,robin07} using
alternative techniques such as decimation~\cite{pasta01}.  

In this paper, we show that the recursive Green function algorithm can be generalized to deal
with arbitrary geometry, topology, number of connected electrodes and inner degrees of freedom
(like spin for ferromagnets or electron/hole in superconductors). Our algorithm is conceptually simpler than the original as it is not based on a specific geometry. The sites are added one by one in a manner reminiscent of the 
knitting of a sweater. The algorithm is optimum in term of speed and a significant gain 
in memory is also achieved. The method allowed us to study the electronic Mach-Zehnder interferometer~\cite{ji03,roull07} 
and anomalous quantum Hall effect in a graphene Hall bar~\cite{novos05} which have been the subject of recent experiments.
Going from the first system to the second required virtually no additional development.
  

\section{NEGF formalism in a nutshell.} 
We consider a quantum system of $N$ sites connected to several conducting electrodes. We use a general tight-biding Hamiltonian
for the system,
\begin{equation}\label{eq:Hamiltonian}
\hat H=\sum_{i\ne j}^Nt_{ij}c_i^\dagger c_j+\sum_{i=1}^N\epsilon_i c_i^\dagger c_i,
\end{equation}
where $c_i^\dagger$  ($c_i$)  denotes the usual creation (annihilation) operator of an electron on site $i$. The site index $i$
stands for the position in space but can also include possible other degrees of freedom like spin 
or electron/hole. $t_{ij}$ is usually very sparse as only nearest (or next nearest) neighboring sites are connected.
The electrodes $l$ are semi-infinite systems at equilibrium with temperature $T_l$ and chemical potential $\mu_l$. They can be ``integrated'' out
of the equations of motion and only appear in the formalism through self-energies $\Sigma_l$ that provide boundary conditions at the connected sites.
 We use an algorithm introduced by Ando~\cite{ando91,khomy05} for the calculation of those self energies. The physical observables can be simply
related to the non equilibrium lesser Green function $G^<_{ij}(E)=i\int dt e^{-i E t}\langle c_j^\dagger c_i(t)\rangle$. For instance the local density $\rho_{i}$ 
and current $I_{ij}$ reads,
\begin{eqnarray}\label{eq:rho_non_equi}
\rho_{i} &=&\frac{1}{2\pi}\text{Im}\int_{-\infty}^{\infty} dE\, G_{ii}^<(E), \\
\label{eq:I_non_equi}
I_{ij} &=&\int_{-\infty}^{\infty} dE\, [t_{ij}G_{ji}^<(E) - t_{ji}G_{ij}^<(E)].
\end{eqnarray}
The simplicity of NEGF comes from the fact that, in absence of electronic correlation, $G^<$ (which describes the system out of equilibrium) 
has a very simple expression in term of the retarded Green function $G$ :
$G^<=G \Sigma^< G^\dagger$ where $\Sigma^<=\sum_l f_l (\Sigma_l^\dagger-\Sigma_l)$ and $f_l=1/(1+\exp [(E-\mu_l)/kT_l])$ is the Fermi function.
$G$ itself is simply defined in term of the one-body Hamiltonian matrix $H_{ij}=t_{ij} + \epsilon_i \delta_{ij}$ by,
\be 
G(E)=1/(E-H-\sum_l \Sigma_l).
\ee

\section{Knitting, sewing and unknitting algorithms}
\subsection{Basic idea behind recursive algorithms.} The problem of
computing $G(E)$ is thus reduced to a conceptually simple task,
finding the inverse of the $H$ matrix (to which we implicitly include the self energies). Our basic tool is to judiciously divide
$H=H_0+V$ between an unperturbed part $H_0$ and a perturbation $V$. Typically, the perturbation will be the hopping elements that allow to
glue two separated part of the system together. The Dyson equation $G=g + g V G$ which
relates $G$ to the known $g=1/(E-H_0)$ is the corner stone of all recursive algorithms. Suppose that (i) one is interested only in $G_{\alpha\beta}$ for 
a small subset of sites labelled by $\alpha,\beta$ (the electrode sites for instance) and (ii) $V_{ij}$ only connects a small number of sites labelled by $i,j$.
The following glueing sequence allows one to get the needed matrix
elements in three steps. (I) Dyson equation restricted to the
connected sites (via $V_{ij}$) is a {\it close} equation: 
\be
G_{ij}=g_{ij}+\sum_{kl} g_{ik} V_{kl} G_{lj}
\ee 
The number of those connected sites being small, the $G_{ij}$ can hence be easily computed. 
(II) In a second step one obtains the elements 
\be 
G_{\alpha j}=g_{\alpha j}+\sum_{kl} g_{\alpha k} V_{kl} G_{lj}
\ee
and (III) in the third step  
\be
G_{\alpha \beta}=g_{\alpha \beta}+\sum_{kl} g_{\alpha k} V_{kl}
G_{l\beta}
\ee
are computed. In the original recursive Green function algorithm~\cite{macki85}, the above sequence is used in the following way: one considers a bar of width $M$ and length $L$. The bar is sliced in $L$ stacks and the perturbation $V_{ij}$ are the hopping elements that connect the different stacks. The system is then built recursively as the stacks are added one at a time and at each step the glueing sequence is used. The $g_{ij}$ are known either from the previous calculation (for the bar side) or from a numerical direct calculation for the added stack.

\subsection{Knitting algorithm for global transport properties.} 
\begin{figure} 
\includegraphics[width=8cm]{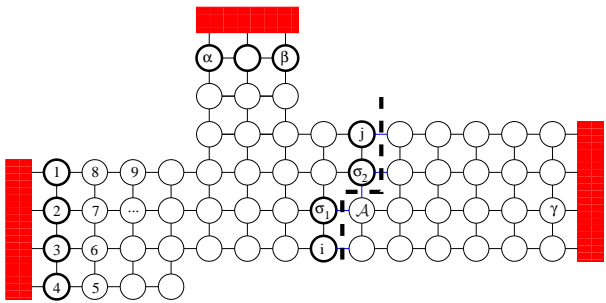}
\caption{\label{fig:schema} The system scheme and notations. The sites are labelled according to the order in which they are added to the system.
The boxes define the three electrodes coupled to the mesoscopic system. The various letters stand for the
added site ($\cal A$), electrode sites ($\alpha,\beta,\gamma$), connected sites ($\sigma_1,\sigma_2$) and interface sites ($i,j$) as discussed in the text. Bold circles indicate sites whose GF elements are updated at current knitting step. The thick dashed line separates the part already included (left) from the part that is still to be knitted to the system (right).}
\end{figure}

Our knitting algorithm is based on the same glueing sequence, but the sites are added one by one. We start by indexing 
the sites according to the order in which they are going to be added to the system. Fig.~\ref{fig:schema} shows a cartoon of a typical system
together with the notations used in the following. The main difficulty of the algorithm lies in the book keeping of the various Green function elements 
and precise definitions are compulsory. At a given stage of the knitting, when we have already included sites $1\dots{\cal A}-1$ and are about to
include site $\cal A$, we distinguish between four categories of sites labelled by different indices: 
\begin{itemize}
\item
the {\it connected sites} are the sites that are directly connected to $\cal A$ via hopping elements $t_{\cal A \sigma}$. They are labelled by the index $\sigma$ and appear in very small number (less then the number of neighbors of a given site $\cal A$).
\item
The {\it interface sites} labelled by $i,j$ are the sites that still miss some of their neighbors, i.e. that will be themselves connected sites later in the knitting. Hence, the Green function elements for those sites is to be kept in memory. Note that it is a dynamical definition, i.e. at each step, some sites appear and others disappear from the interface. The total number $M$ of interface sites scales as a surface $M\propto L^{d-1}$. 
\item
{\it Updated sites}, the sites whose Green function elements are updated at each step of knitting. They belong either to the interface or to the electrodes. They are noted by $\alpha,\beta$; bold circles on Fig.~\ref{fig:schema}.
\end{itemize}    
With these notations, we can apply the glueing sequence and express the Green function with the added site $\GA$ in term of the Green function $\GAm$ of the system composed of ${\cal A} -1$ sites. (I) The first step reads, 
\begin{equation}
\label{eq:knitI}
\GA_{\cal AA}=1/(E-\epsilon_{\cal A}-\sum_{\sigma\tilde\sigma}t_{\cal A\sigma}\GAm_{\sigma\tilde\sigma}t_{\tilde\sigma\cal A})
\end{equation}
Note that it is the only place where we actually perform an inversion and it is done on a scalar quantity. (II) The second step reads,
\begin{eqnarray}
\GA_{\alpha\cal A}&=&\sum\nolimits_{\sigma} \GAm_{\alpha\sigma}t_{\sigma\cal A}\GA_{\cal AA}, \label{eq:fwG_aA}\\
\GA_{\cal A\beta}&=&\sum\nolimits_{\sigma} \GA_{\cal AA}t_{\cal A\sigma}\GAm_{\sigma\beta}, \label{eq:fwG_Aa}
\end{eqnarray}
(III) The last step concentrates almost all the computing time ($\propto M^2$), 
\begin{eqnarray}
\label{eq:knitIII}
\GA_{\alpha\beta}&=&\GAm_{\alpha\beta}+\GA_{\alpha\cal A} \frac{1}{\GA_{\cal AA}}\GA_{\cal A\beta}, \label{eq:fwG_ab}
\end{eqnarray}
Note that the previous formula has a very simple physical interpretation in term of paths: 
the amplitude for an electron to go from site $\beta$ to site $\alpha$ is the amplitude which avoids site $\cal A$ 
plus the amplitude which goes through site $\cal A$. The factor $1/\GA_{\cal AA}$ removes the double counting of the loops from $\cal A$ to itself. Once the glueing sequence is completed, the interface is updated: the new site is added while sites that now have all their neighbors can be removed. For instance in Fig.~\ref{fig:schema}, once site $\cal A$ has been added, site $\sigma_1$ can be deleted from the interface. The previous sequence is iterated until all the $N$ sites have been added to the system. Eventually we get $G_{\alpha\beta}$ of the entire system. 
However, we only get the matrix elements between the (few) sites $\alpha,\beta$ connected to the electrodes. Those matrix elements give access to 
all transport properties like conductance or shot noise, but no information on what happens inside the sample. The computing time (needed memory) of the knitting algorithm scales as $M^2 N$ ($M^2$) in agreement with the original recursive algorithm for a wire.

\subsection{A sewing algorithm for calculating local observables.} We now proceed with extending the knitting 
algorithm to the calculations of local observables. In practice, we need matrix elements of the type
$G_{\alpha r}$ between electrodes and any inner site of the system $r$. Such an extension has been done in \cite{metal05b}
for the original recursive algorithm: one performs a first recursive calculation and saves the partial
Green functions $G_{\alpha r}^{[r]}$ for future use. When the calculation is completed, one starts a new
recursive calculation, beginning from {\it the other} hand of the system. Along the way, one recovers the saved Green functions for the left part of the system and uses the glueing sequence to glue them
with the (freshly calculated) Green function of the right part of the system. This scheme can be generalized to an arbitrary geometry as well: we first perform a full knitting calculation. Then we start backward the sewing algorithm and sew the sites $\cal A$ one by one in reversed order (i.e. from $N$ to $1$). We label sites with an index 
smaller than $\cal A$ ("left" part of the system) by indices without prime ($i,j,...$) and the sites (strictly) higher than $\cal A$ ("right") by prime indices ($i',j',...$). Introducing the "interface self energy" ${\mathbb S}_{ij}=\sum_{i'j'}t_{ij'}G_{j'i'}t_{i'j}$, the glueing sequence
reads,
\begin{eqnarray}
G_{\alpha\cal A}&=&\GA_{\alpha\cal A}+\sum_{ij}\GA_{\alpha i}{\mathbb S}_{ij}\GA_{j\cal A},\quad \\
G_{\alpha'\cal A}&=&\sum_{j'i}G_{\alpha'j'}t_{j'i}\GA_{i\cal A}.
\end{eqnarray}
The result of the calculation is then used to update ${\mathbb S}_{ij}$ and one can proceed with ${\cal A}-1$ and so on. The drawback of this algorithm is that many ($\sim N M$) matrix elements
$G^{[r]}_{\alpha r}$ must be stored during the first knitting calculation which limit practical calculations to a few hundred thousand sites.

\subsection{Saving memory with unknitting.} The last piece of algorithm, called
unknitting, allows to recalculate the matrix elements $\GA_{\alpha\cal A}$ in the 
backward calculation instead of saving them. Indeed, using Dyson equation for $H_0=H-V$,
it is possible to "remove" sites from the system and express $\GAm$ as a function of $\GA$ 
\begin{equation}
\label{eq:bwG_ab}
\GAm_{\alpha\beta}=\GA_{\alpha\beta}-\GA_{\alpha\cal A} \frac{1}{\GA_{\cal AA}}\GA_{\cal A\beta}.
\end{equation}
The above equation should however be taken with care. Indeed the interface is not the same
for $\GAm$ and $\GA$ so that some of the matrix elements $\GA_{\alpha\cal A}$ on the right hand side are not stored in memory. In the bulk of the system the interface is of constant size so that
there is only one site $\cal R$ that belongs to the interface of $\GAm$ but not of $\GA$. The matrix elements for this site can also be recomputed,  
\begin{eqnarray}
\GAm_{\alpha\cal R}&=&\GA_{\alpha\cal A}(t_{\cal RA}G_{\cal AA})^{-1}
-\sum_{\sigma\ne\cal R} \GAm_{\alpha\sigma}t_{\sigma\cal A}t_{\cal RA}^{-1}\,\nonumber \\
\GAm_{\cal R\alpha}&=&(\GA_{\cal AA}t_{\cal AR})^{-1}\GA_{\cal A\alpha} 
-\sum_{\sigma\ne\cal R} t_{\cal AR}^{-1} t_{\cal A\sigma}\GAm_{\sigma\alpha}\nonumber
\end{eqnarray}
which completes the algorithm. Theoretically, the unknitting algorithm allows to decrease the memory from $M N$ to 
$max (N,M^2)$, hence completely removing the memory bottleneck in the calculations. We found that it is indeed the case
in the middle of the tight-biding band where all the channels are conducting. Outside this region, the last equation above is however numerically unstable and introduces an error in the calculation that increases as $\exp (a N/M)$. The origin of this instability can be understood from the example of a simple perfect 1D chain with unit hopping. 
In this case, Eq.(\ref{eq:knitI}) takes a simple form $\GA=1/(E-\GAm)$. When $|E|>2$, i.e. the chain is 
no longer propagative (only evanescent waves can be found) and this equation converges toward a simple 
attractive fixed point. After several
iterations, the convergence is achieved up to numerical precision and the initial condition is lost: the equation can no longer be inverted to follow our steps back. We found that the unknitting algorithm is nevertheless useful. Depending on the precision needed, the matrix elements $\GA_{\alpha\cal A}$ can be saved in the forward calculation only every few sites instead of every site. In practice, in the worth case (bottom of the band where almost all modes are evanescent), 
we found that a factor $10$ in memory could be gained while keeping the numerical precision better than $10^{-10}$.


\section{ Application: Mach-Zehnder interferometer in a two dimensional electron gas.}
\begin{figure}
\includegraphics[width=8cm]{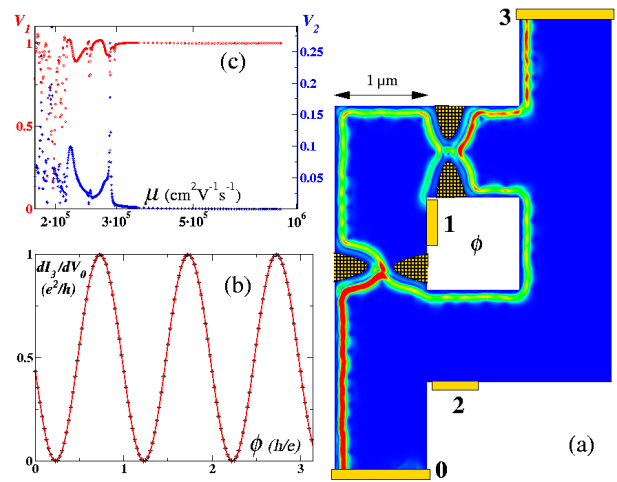}
\caption{\label{fig:machzehnder} Mach-Zehnder interferometer in a 2D gas modelled by a scalar 
tight-biding model. Mobility is $\mu=5\cdot10^5\rm{cm}^2/\text{Vs}$, electron density $n_s=10^{10}\text{cm}^{-2}$ and
 magnetic field $B=0.2$ T. (a): local current intensity when a bias voltage is applied to lead 0 and the other contacts are grounded ($1.2$ million sites, blue colors corresponds to no current and red to maximum current). (b): differential conductance $dI_3/dV_0$ as a function of the number of flux quanta $\phi$ through the hole.  
(c): visibility coefficients ${\cal V}_1$ and ${\cal V}_2$ (see text) as a function of the electron mobility $\mu$.}
\end{figure}
 As a first application of our algorithm, we consider a Mach-Zehnder interferometer similar to the ones studied in recent 
experiments~\cite{roull07}. The devices are made in a two-dimensional high mobility GaAs/AlGaAs heterostructures
which we characterize by its mobility $\mu$, electronic density $n_s$, size $\cal L$ and perpendicular magnetic field $B$.
The simulations are performed by discretizing the Schr\"odinger equation on a square grid with a lattice spacing $b$.
The resulting tight-biding model has nearest neighbor hoppings $t$ and disordered on-site potential which are chosen
randomly within $[-W/2,+W/2]$. The magnetic field is added within a discretized Landau gauge by adding a phase $\Phi$ in the
hoppings elements  $t_{(n_x,n_y)}^y$ along the $y$ direction: $t_{(n_x,n_y)}^y = t e^{i2\pi\Phi n_x}$ 
(where $(n_x,n_y)$ is the position on the grid along $x$ and $y$). 
The energy $E$ is measured from the bottom of the band, and we found no significant
deviations from the continuum limit as long as we kept $E <0.1 t$. 
The tight-biding parameters are related to the experimental ones as follows: $n_s= E/(2\pi t b^2)$, $B=\Phi h/(e b^2)$ and $\mu=(e/h) 96\pi b^2 (t/W)^2$. The latter formula was obtained from a calculation of the Drude conductance
of our tight-biding model and was checked by direct numerical calculations. The total number ${\cal N = L}^2 n_s$ of electrons
in the sample and the conductance per square $g=e\mu n_s$ are both independent of the discretized step $b$.
The middle of the $n^{th}$ plateau of quantum Hall effect is found for $n=E/(4\pi t \Phi)$.

\begin{figure}
\includegraphics[width=8cm]{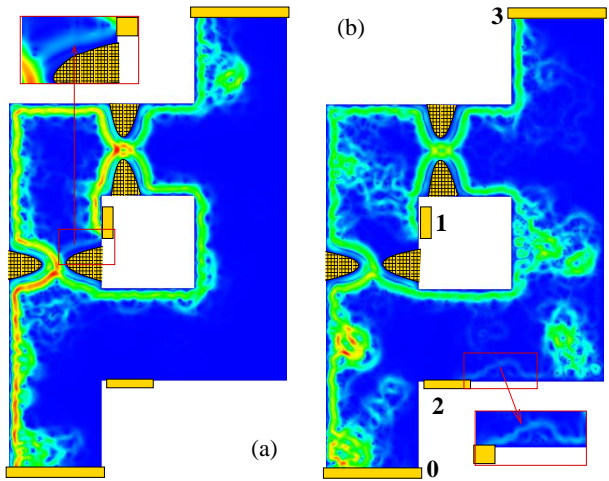}
\caption{\label{fig:machzehnderFlux} Same as Fig.\ref{fig:machzehnder}a for two dirtier samples with mobility $\mu=3\cdot10^5\rm{cm}^2/\text{Vs}$ (a) and $\mu=2.5\cdot10^5\rm{cm}^2/\text{Vs}$ (b), electron density $n_s=10^{10}\text{cm}^{-2}$ and magnetic field $B=0.2$ T. Zoom of (a) shows the current that escapes the first contact, and makes the second loop. It is responsible for the presence of a second visibility harmonic ${\cal V}_1=1$ ${\cal V}_2=0.1$. Similar zoom of (b) shows backscattered current which is recovered by contact~2. ${\cal V}_1=0.99$ ${\cal V}_2=0.01$.}
\end{figure}
The Mach-Zehnder interferometer is the electronic analog of the well-known optical device. The 
system consists of a loop connected to four contacts, one of which lies in the center part of 
the loop, see Fig.~\ref{fig:machzehnder}a. The physics involved is fairly 
straightforward: the system is placed under a high magnetic field in the quantum Hall regime 
at the first Hall plateau (the current is supported by one edge state). All contacts are grounded except 
contact $0$ which is placed at a slightly higher voltage $V_0$. The injected current follows the edge channel until it
reaches a first QPC (Quantum Point Contact which works as a perfect beam splitter) and is split into
two parts, see Fig.~\ref{fig:machzehnder}a. The two edge states are eventually recombined at the second
QPC and the current $I_3$ is collected in contact $3$. Along the way, the two edge states pick up a difference of phases that
includes the magnetic flux $\phi$ through the hole. 
Note that Fig.~\ref{fig:machzehnder}a is not a cartoon, but an actual calculation of the local density of current injected from lead $0$. Along the way, the two edge channels pick up a phase difference which produces interferences. Fig.~\ref{fig:machzehnder}b shows the differential conductance $dI_3/dV_0$ in unit of $e^2/h$ as a function of flux $\phi$ (in unit of $h/e$) along with a (almost perfect) sinusoidal fit. Our data are in complete agreement with what is obtained from Landauer-B\"uttiker theory. In particular, the visibility ${\cal V}_1$ decreases when the transmission $T$ of one QPC departs from $1/2$ as ${\cal V}_1= 2\sqrt{T(1-T)}$. 

To proceed further, we increase the disorder in our sample and measure the visibility of the interference pattern as a function of the mobility $\mu$. We find that  $dI_3/dV_0(\phi)$ is well fitted by including two harmonics, 
\be
dI_3/dV_0\propto  1 + {\cal V}_1 \cos 2\pi\phi +{\cal V}_2 \cos 4\pi\phi
\ee
The parameters ${\cal V}_1$ and ${\cal V}_2$ are shown in  Fig.~\ref{fig:machzehnder}c for a typical sample. 
For $\mu\ge 4\cdot 10^5 {\rm cm^2 V^{-1} s^{-1}}$, there is no backscattering in the sample 
and the interference pattern is perfect ${\cal V}_1=1$ and ${\cal V}_2=0$. Below  $4\cdot 10^5 {\rm cm^2 V^{-1} s^{-1}}$
however, backscattering sets in and the visibility decreases strongly. More importantly, the visibility becomes
extremely sensitive to the disorder configuration and huge sample to sample fluctuations are observed. Simultaneously,
the second harmonic ${\cal V}_2$ sets up. Those results are readily understood by looking at the current intensities
injected from lead $0$, as shown in Fig.~\ref{fig:machzehnderFlux} for two different samples with mobilities slightly lower then in Fig.~\ref{fig:machzehnder}a. When the disorder is strong enough, the edge channels get significantly disturbed and are able to reach (partially) the opposite wall (as seen in Fig.~\ref{fig:machzehnderFlux}b). This backscattering leads to a decrease
of the visibility. At the same time, a path that avoids contact $1$ appears which causes ${\cal V}_2$ to increase (the corresponding path makes a complete tour of the central island, see the inset of Fig.~\ref{fig:machzehnderFlux}a).  It is notable that no second harmonic ${\cal V}_2$ was observed
experimentally which is, as the samples had mobility of the order of
$10^6 {\rm cm^2 V^{-1} s^{-1}}$ 
or higher~\cite{ji03,roull07}, consistent with our findings.
We conclude that static disorder (or more trivially, a bad central ohmic contact)
cannot be at the origin of the reduced visibility in these experiments.

\section{Application: anomalous quantum Hall effect in a graphene Hall
  bar.}

\begin{figure}
\includegraphics[width=8cm]{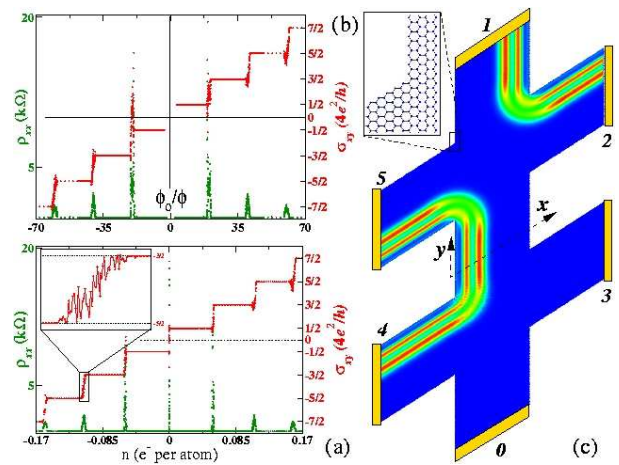}
\caption{\label{fig:graphene} Quantum Hall effect in graphene. Hall 
conductance $\sigma_{xy}$ and longitudinal resistance $\rho_{xx}$ are plotted as a function of 
inverse magnetic field $1/\Phi$ (a) and carrier density $n$ (b) in presence of a small disordered potential 
(10\% of the hopping matrix elements). In (a) carrier density is $0.18$ electrons per atom while in (b) a magnetic 
flux $\Phi=0.014h/e$ per hexagon is applied.
The inset of (b) shows a zoom of the transition between plateaus. (c) shows
the local current intensity when current is injected from both contact $1$ and $4$ 
($3\cdot10^5$ hexagons, $N=2$)}
\end{figure}

In this second application, we consider a Hall bar in graphene in 
the quantum Hall regime, see Fig.~\ref{fig:graphene}c. 
We used the standard tight binding approach~\cite{neto07}  for the graphene hexagonal lattice, where 
each site represents a carbon atom and is connected to his neighbors by a hopping $t$. We use zigzag edges for
the system and the leads which are semi infinite nanoribbons. To avoid reflection at the lead/system interface,
the leads also include the magnetic field. The magnetic field is included within a gauge similar to the one
used for the Mach-Zehnder above, by introducing a phase $\Phi$ in the hopping elements between the A and B sites.
Noting $a$ (1.42\AA)  the distance between carbon atoms, the electronic density $n_s$ and magnetic field $B$ are related
to $E/t$ and the total flux $\Phi$ (in unit of $h/e$) per hexagon as follows: $n_s=(E/at)^2 8/(9\pi)$ and
$B=2h\Phi/(3\sqrt{3}e a^2)$. The middle of $N^{th}$ Hall plateau is found for $N+1/2=E^2/(2\sqrt{3}\pi\Phi)$

Fig.~\ref{fig:graphene} shows the transverse conductance and longitudinal resistance as
a function of carrier density (a) or inverse magnetic flux (b) for a graphene Hall bar.  
Our result for a mesoscopic Hall bar are numerical counterparts to the experimental data of~\cite{novos05}. 
In particular we recover the presence of plateaus in the transverse conductance at quantized values $(N+1/2)\cdot4e^2/h$. The $N=0$ plateau corresponds to a Landau level pinned to the Dirac point, and is special to the Dirac equation symmetry class. We also observe mesoscopic fluctuations at the transition between plateaus similar to those known in usual 2D
gases, see inset of Fig.~\ref{fig:graphene}a. The edge current density at $N=2$ injected
simultaneously from contact 1 and 4 is plotted in Fig.~\ref{fig:graphene}c which is the graphene counterpart to
Fig.~\ref{fig:machzehnder}c.

\begin{figure}
\includegraphics[width=8cm]{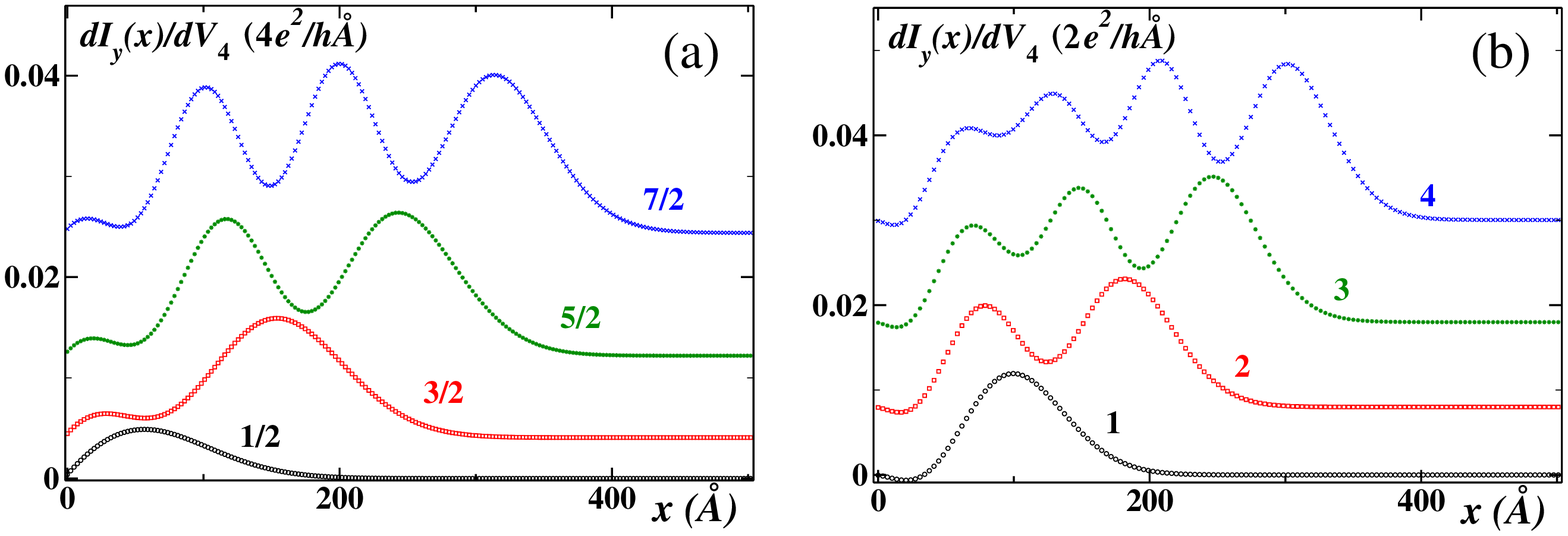}
\caption{\label{fig:edge_profile} Comparison of the edge channels for a graphene (a) and a semi-conductor
heterostructure (b) two-dimensional electron gases. The plots show the local current density in the $y$ direction 
$dI_y(x)/dV$ as a function of the distance $x$ from the border of the sample for four consecutive Hall plateaus corresponding to $N=0,1,2,3$ for graphene (a) and $n=1,2,3,4$  for the heterostructure (b). All graphs are translated for clarity from bottom to top. Magnetic field $B=14$ T.}
\end{figure}

We note one important difference between the edge states of graphene and those of a regular heterostructure: 
while Fig.~\ref{fig:graphene}c has been made for $N=2$, i.e. for the third plateau in the sequence, one observes 
only two peaks in the local current density, i.e. apparently two edge states only.
This is revealed more quantitatively in Fig.\ref{fig:edge_profile} where we have plotted the cross section of the current density for the first plateaus of both the graphene ($N=0,1,2,3$) and the heterostructure ($n=1,2,3,4$). The area under
those curves correspond respectively to $1/2$, $3/2$, $5/2$ and $7/2$ in the graphene and $1$, $2$, $3$ and $4$ in the 
heterostructure. The number of peaks however is respectively $1$, $1$, $2$ and $3$ and $1$, $2$, $3$ and $4$ so that the 
$N=0$ peak (corresponding to the $E=0$ Landau level) gets blurred behind the other peaks upon increasing carrier
density. This peculiar behavior could possibly be observed in STM or local compressibility experiments~\cite{zhite05}.

\section{Conclusion}

We have presented a set of algorithms that allows to calculate the transport and local
properties of a generic class of tight-biding models. Using the unknitting technique, some
significant gain of memory could be gained with respect to previous techniques. However, our main point
here is the global simplicity of the algorithm. We could easily implement it to get a versatile code that 
treats complicated geometries, such as the ones presented in the applications, with the same ease as one
treats the usual quasi-one dimensional bar, ubiquitous in the literature.

\appendix
\section{Implementation tips.}

In this appendix, we have grouped a few technical points which
can help to get significant gains in computing time, or maybe more 
importantly in human development time.

1. The computing time for adding one site in the knitting algorithm is
dominated by the last step of the glueing sequence, Eq.(\ref{eq:knitIII}) and
scales as $O(M^2)$. Hence, almost all the computing time of the algorithm is concentrated
in a single line of code which can be optimized aggressively.
The best results were obtained by using low level BLAS routines for this simple 
step. Note that the only inversions that take place in the algorithm are scalars: at no point
we need to rely on matrix inversion routines in contrast to the original recursive technique.

2. A practical difficulty is to maintain dynamically the list of active {\it interface} sites.
We use the following simple algorithm: to each {\it interface} site is associated a list of the site's
neighbors that have not yet been knitted to the system. As new sites are knitted to the system, this list 
is updated. When a site has all his neighbors knitted, he can be removed from the interface.

3. The algorithms described in this paper allow to solve a very generic class of tight-biding models.
Almost as important as those solvers is the ability to easily construct new systems. In our implementation,
we represent internally a system as a generic graph: each site possesses the list of its neighbors, together
with the corresponding matrix elements. A set of routines is used to operate on those graphes. A very important 
one is a function that allows to stick two systems together by identifying some sites of the two systems. This routine
allows us to construct a system from small pieces in a ``Lego'' like way. For instance, the system of 
Fig.~\ref{fig:machzehnder}c is obtained in four lines of code by ``sticking'' four rectangular systems together.

4. The intensive pieces of the code should be written in a low level language (in our case C++). However,
we have found that very significant gain in developing time could be obtained by making use of the code in
a higher level scripting language. In our case, we use Python, and the automatic wrapping of the C++ code
into the Python API has been done with SWIG. The systems constructions, calculations and plotting are done in 
small Python programs that constitute our input files.

\acknowledgments
We wish to thank O. Parcollet, A. Freyn and F. Portier for useful discussions.  
Financial support from region "Ile de France" as well as from the
EC Contracts IST-033749 "DynaMax" is acknowledged.


\bibliographystyle{apsrev}
\bibliography{knit}

\begin{thebibliography}{31}
\expandafter\ifx\csname natexlab\endcsname\relax\def\natexlab#1{#1}\fi
\expandafter\ifx\csname bibnamefont\endcsname\relax
  \def\bibnamefont#1{#1}\fi
\expandafter\ifx\csname bibfnamefont\endcsname\relax
  \def\bibfnamefont#1{#1}\fi
\expandafter\ifx\csname citenamefont\endcsname\relax
  \def\citenamefont#1{#1}\fi
\expandafter\ifx\csname url\endcsname\relax
  \def\url#1{\texttt{#1}}\fi
\expandafter\ifx\csname urlprefix\endcsname\relax\def\urlprefix{URL }\fi
\providecommand{\bibinfo}[2]{#2}
\providecommand{\eprint}[2][]{\url{#2}}

\bibitem[{\citenamefont{Metalidis and Bruno}(2005)}]{metal05b}
\bibinfo{author}{\bibfnamefont{G.}~\bibnamefont{Metalidis}} \bibnamefont{and}
  \bibinfo{author}{\bibfnamefont{P.}~\bibnamefont{Bruno}},
  \bibinfo{journal}{Phys.~Rev.~B} \textbf{\bibinfo{volume}{72}},
  \bibinfo{eid}{235304} (\bibinfo{year}{2005}).

\bibitem[{\citenamefont{Rycerz and Beenakker}(2007)}]{rycer07}
\bibinfo{author}{\bibfnamefont{A.}~\bibnamefont{Rycerz}} \bibnamefont{and}
  \bibinfo{author}{\bibfnamefont{C.~W.~J.} \bibnamefont{Beenakker}},
  \bibinfo{journal}{cond-mat/0709.3397}  (\bibinfo{year}{2007}).

\bibitem[{\citenamefont{Gunsenheimer and Zaikin}(1994)}]{gunse94}
\bibinfo{author}{\bibfnamefont{U.}~\bibnamefont{Gunsenheimer}}
  \bibnamefont{and} \bibinfo{author}{\bibfnamefont{A.~D.}
  \bibnamefont{Zaikin}}, \bibinfo{journal}{Phys.~Rev.~B}
  \textbf{\bibinfo{volume}{50}}, \bibinfo{pages}{6317} (\bibinfo{year}{1994}).

\bibitem[{\citenamefont{Damle et~al.}(2001)\citenamefont{Damle, Ghosh, and
  Datta}}]{damle01}
\bibinfo{author}{\bibfnamefont{P.~S.} \bibnamefont{Damle}},
  \bibinfo{author}{\bibfnamefont{A.~W.} \bibnamefont{Ghosh}}, \bibnamefont{and}
  \bibinfo{author}{\bibfnamefont{S.}~\bibnamefont{Datta}},
  \bibinfo{journal}{Phys.~Rev.~B} \textbf{\bibinfo{volume}{64}},
  \bibinfo{pages}{201403} (\bibinfo{year}{2001}).

\bibitem[{\citenamefont{Sanvito et~al.}(1999)\citenamefont{Sanvito, Lambert,
  Jefferson, and Bratkovsky}}]{sanvi99}
\bibinfo{author}{\bibfnamefont{S.}~\bibnamefont{Sanvito}},
  \bibinfo{author}{\bibfnamefont{C.~J.} \bibnamefont{Lambert}},
  \bibinfo{author}{\bibfnamefont{J.~H.} \bibnamefont{Jefferson}},
  \bibnamefont{and} \bibinfo{author}{\bibfnamefont{A.~M.}
  \bibnamefont{Bratkovsky}}, \bibinfo{journal}{Phys. Rev. B}
  \textbf{\bibinfo{volume}{59}}, \bibinfo{pages}{11936} (\bibinfo{year}{1999}).

\bibitem[{\citenamefont{Haney et~al.}(2007)\citenamefont{Haney, Waldron, Duine,
  Nunez, Guo, and MacDonald}}]{haney07}
\bibinfo{author}{\bibfnamefont{P.~M.} \bibnamefont{Haney}},
  \bibinfo{author}{\bibfnamefont{D.}~\bibnamefont{Waldron}},
  \bibinfo{author}{\bibfnamefont{R.~A.} \bibnamefont{Duine}},
  \bibinfo{author}{\bibfnamefont{A.~S.} \bibnamefont{Nunez}},
  \bibinfo{author}{\bibfnamefont{H.}~\bibnamefont{Guo}}, \bibnamefont{and}
  \bibinfo{author}{\bibfnamefont{A.~H.} \bibnamefont{MacDonald}},
  \bibinfo{journal}{Phys.~Rev.~B} \textbf{\bibinfo{volume}{76}},
  \bibinfo{eid}{024404} (\bibinfo{year}{2007}).

\bibitem[{\citenamefont{Blanter and Buttiker}(2000)}]{blant00}
\bibinfo{author}{\bibfnamefont{Y.~M.} \bibnamefont{Blanter}} \bibnamefont{and}
  \bibinfo{author}{\bibfnamefont{M.}~\bibnamefont{Buttiker}},
  \bibinfo{journal}{Phys.~Rep.~} \textbf{\bibinfo{volume}{336}},
  \bibinfo{pages}{1} (\bibinfo{year}{2000}).

\bibitem[{\citenamefont{C~Caroli and Saint-James}(1971)}]{carol71}
\bibinfo{author}{\bibfnamefont{P.~N.} \bibnamefont{C~Caroli},
  \bibfnamefont{R~Combescot}} \bibnamefont{and}
  \bibinfo{author}{\bibfnamefont{D.}~\bibnamefont{Saint-James}},
  \bibinfo{journal}{J.~Phys.~C} \textbf{\bibinfo{volume}{4}},
  \bibinfo{pages}{916} (\bibinfo{year}{1971}).

\bibitem[{\citenamefont{Meir and Wingreen}(1992)}]{meir92}
\bibinfo{author}{\bibfnamefont{Y.}~\bibnamefont{Meir}} \bibnamefont{and}
  \bibinfo{author}{\bibfnamefont{N.~S.} \bibnamefont{Wingreen}},
  \bibinfo{journal}{Phys.~Rev.~Lett.} \textbf{\bibinfo{volume}{68}},
  \bibinfo{pages}{2512} (\bibinfo{year}{1992}).

\bibitem[{\citenamefont{Keldysh}(1964)}]{keldy64}
\bibinfo{author}{\bibfnamefont{L.}~\bibnamefont{Keldysh}},
  \bibinfo{journal}{Zh.~Eksp.~Teor.~Fiz.} \textbf{\bibinfo{volume}{47}},
  \bibinfo{pages}{1515} (\bibinfo{year}{1964}).

\bibitem[{\citenamefont{Lake and Pandey}(2006)}]{lake06}
\bibinfo{author}{\bibfnamefont{R.~K.} \bibnamefont{Lake}} \bibnamefont{and}
  \bibinfo{author}{\bibfnamefont{R.~R.} \bibnamefont{Pandey}},
  \bibinfo{journal}{cond-mat/0607219}  (\bibinfo{year}{2006}).

\bibitem[{\citenamefont{Brandbyge et~al.}(2002)\citenamefont{Brandbyge, Mozos,
  Ordej\'on, Taylor, and Stokbro}}]{brand02}
\bibinfo{author}{\bibfnamefont{M.}~\bibnamefont{Brandbyge}},
  \bibinfo{author}{\bibfnamefont{J.-L.} \bibnamefont{Mozos}},
  \bibinfo{author}{\bibfnamefont{P.}~\bibnamefont{Ordej\'on}},
  \bibinfo{author}{\bibfnamefont{J.}~\bibnamefont{Taylor}}, \bibnamefont{and}
  \bibinfo{author}{\bibfnamefont{K.}~\bibnamefont{Stokbro}},
  \bibinfo{journal}{Phys.~Rev.~B} \textbf{\bibinfo{volume}{65}},
  \bibinfo{pages}{165401} (\bibinfo{year}{2002}).

\bibitem[{\citenamefont{Khomyakov et~al.}(2005)\citenamefont{Khomyakov, Brocks,
  Karpan, Zwierzycki, and Kelly}}]{khomy05}
\bibinfo{author}{\bibfnamefont{P.~A.} \bibnamefont{Khomyakov}},
  \bibinfo{author}{\bibfnamefont{G.}~\bibnamefont{Brocks}},
  \bibinfo{author}{\bibfnamefont{V.}~\bibnamefont{Karpan}},
  \bibinfo{author}{\bibfnamefont{M.}~\bibnamefont{Zwierzycki}},
  \bibnamefont{and} \bibinfo{author}{\bibfnamefont{P.~J.} \bibnamefont{Kelly}},
  \bibinfo{journal}{Phys.~Rev.~B} \textbf{\bibinfo{volume}{72}},
  \bibinfo{eid}{035450} (pages~\bibinfo{numpages}{13}) (\bibinfo{year}{2005}).

\bibitem[{\citenamefont{MacKinnon}(1985)}]{macki85}
\bibinfo{author}{\bibfnamefont{A.}~\bibnamefont{MacKinnon}},
  \bibinfo{journal}{Zeit.~f.~Phys.~B} \textbf{\bibinfo{volume}{59}},
  \bibinfo{pages}{385} (\bibinfo{year}{1985}).

\bibitem[{\citenamefont{Thouless and Kirkpatrick}(1981)}]{thoul81}
\bibinfo{author}{\bibfnamefont{D.~J.} \bibnamefont{Thouless}} \bibnamefont{and}
  \bibinfo{author}{\bibfnamefont{S.}~\bibnamefont{Kirkpatrick}},
  \bibinfo{journal}{J.~Phys.~C} \textbf{\bibinfo{volume}{14}},
  \bibinfo{pages}{235} (\bibinfo{year}{1981}).

\bibitem[{\citenamefont{Lee and Fisher}(1981)}]{lee81}
\bibinfo{author}{\bibfnamefont{P.~A.} \bibnamefont{Lee}} \bibnamefont{and}
  \bibinfo{author}{\bibfnamefont{D.~S.} \bibnamefont{Fisher}},
  \bibinfo{journal}{Phys. Rev. Lett.} \textbf{\bibinfo{volume}{47}},
  \bibinfo{pages}{882} (\bibinfo{year}{1981}).

\bibitem[{\citenamefont{Cresti et~al.}(2003)\citenamefont{Cresti, Farchioni,
  Grosso, and Parravicini}}]{crest03}
\bibinfo{author}{\bibfnamefont{A.}~\bibnamefont{Cresti}},
  \bibinfo{author}{\bibfnamefont{R.}~\bibnamefont{Farchioni}},
  \bibinfo{author}{\bibfnamefont{G.}~\bibnamefont{Grosso}}, \bibnamefont{and}
  \bibinfo{author}{\bibfnamefont{G.~P.} \bibnamefont{Parravicini}},
  \bibinfo{journal}{Phys.~Rev.~B} \textbf{\bibinfo{volume}{68}},
  \bibinfo{pages}{075306} (\bibinfo{year}{2003}).

\bibitem[{\citenamefont{Baranger et~al.}(1991)\citenamefont{Baranger,
  DiVincenzo, Jalabert, and Stone}}]{baran91}
\bibinfo{author}{\bibfnamefont{H.~U.} \bibnamefont{Baranger}},
  \bibinfo{author}{\bibfnamefont{D.~P.} \bibnamefont{DiVincenzo}},
  \bibinfo{author}{\bibfnamefont{R.~A.} \bibnamefont{Jalabert}},
  \bibnamefont{and} \bibinfo{author}{\bibfnamefont{A.~D.} \bibnamefont{Stone}},
  \bibinfo{journal}{Phys.~Rev.~B} \textbf{\bibinfo{volume}{44}},
  \bibinfo{pages}{10637} (\bibinfo{year}{1991}).

\bibitem[{\citenamefont{Rotter et~al.}(2003)\citenamefont{Rotter, Weingartner,
  Rohringer, and Burgd\"orfer}}]{rotte03}
\bibinfo{author}{\bibfnamefont{S.}~\bibnamefont{Rotter}},
  \bibinfo{author}{\bibfnamefont{B.}~\bibnamefont{Weingartner}},
  \bibinfo{author}{\bibfnamefont{N.}~\bibnamefont{Rohringer}},
  \bibnamefont{and}
  \bibinfo{author}{\bibfnamefont{J.}~\bibnamefont{Burgd\"orfer}},
  \bibinfo{journal}{Phys. Rev. B} \textbf{\bibinfo{volume}{68}},
  \bibinfo{pages}{165302} (\bibinfo{year}{2003}).

\bibitem[{\citenamefont{Sols et~al.}(1989)\citenamefont{Sols, Macucci,
  Ravaioli, and Hess}}]{sols89}
\bibinfo{author}{\bibfnamefont{F.}~\bibnamefont{Sols}},
  \bibinfo{author}{\bibfnamefont{M.}~\bibnamefont{Macucci}},
  \bibinfo{author}{\bibfnamefont{U.}~\bibnamefont{Ravaioli}}, \bibnamefont{and}
  \bibinfo{author}{\bibfnamefont{K.}~\bibnamefont{Hess}}, \bibinfo{journal}{J.~
  Appl.~Phys.} \textbf{\bibinfo{volume}{66}}, \bibinfo{pages}{3892}
  (\bibinfo{year}{1989}).

\bibitem[{\citenamefont{Polin\'{a}k et~al.}(2006)\citenamefont{Polin\'{a}k,
  Lambert, Koltai, and Cserti}}]{polin06}
\bibinfo{author}{\bibfnamefont{P.~K.} \bibnamefont{Polin\'{a}k}},
  \bibinfo{author}{\bibfnamefont{C.~J.} \bibnamefont{Lambert}},
  \bibinfo{author}{\bibfnamefont{J.}~\bibnamefont{Koltai}}, \bibnamefont{and}
  \bibinfo{author}{\bibfnamefont{J.}~\bibnamefont{Cserti}},
  \bibinfo{journal}{Phys.~Rev.~B} \textbf{\bibinfo{volume}{74}},
  \bibinfo{eid}{132508} (\bibinfo{year}{2006}).

\bibitem[{\citenamefont{Guan et~al.}(2003)\citenamefont{Guan, Ravaioli,
  Giannetta, Hannan, Adesida, and Melloch}}]{guan03}
\bibinfo{author}{\bibfnamefont{D.}~\bibnamefont{Guan}},
  \bibinfo{author}{\bibfnamefont{U.}~\bibnamefont{Ravaioli}},
  \bibinfo{author}{\bibfnamefont{R.~W.} \bibnamefont{Giannetta}},
  \bibinfo{author}{\bibfnamefont{M.}~\bibnamefont{Hannan}},
  \bibinfo{author}{\bibfnamefont{I.}~\bibnamefont{Adesida}}, \bibnamefont{and}
  \bibinfo{author}{\bibfnamefont{M.~R.} \bibnamefont{Melloch}},
  \bibinfo{journal}{Phys. Rev. B} \textbf{\bibinfo{volume}{67}},
  \bibinfo{pages}{205328} (\bibinfo{year}{2003}).

\bibitem[{\citenamefont{Robinson and Schomerus}(2007)}]{robin07}
\bibinfo{author}{\bibfnamefont{J.~P.} \bibnamefont{Robinson}} \bibnamefont{and}
  \bibinfo{author}{\bibfnamefont{H.}~\bibnamefont{Schomerus}},
  \bibinfo{journal}{Phys.~Rev.~B} \textbf{\bibinfo{volume}{76}},
  \bibinfo{eid}{115430} (\bibinfo{year}{2007}).

\bibitem[{\citenamefont{Wimmer et~al.}(2007)\citenamefont{Wimmer, Adagideli,
  Berber, Tomanek, and Richter}}]{wimme07}
\bibinfo{author}{\bibfnamefont{M.}~\bibnamefont{Wimmer}},
  \bibinfo{author}{\bibfnamefont{I.}~\bibnamefont{Adagideli}},
  \bibinfo{author}{\bibfnamefont{S.}~\bibnamefont{Berber}},
  \bibinfo{author}{\bibfnamefont{D.}~\bibnamefont{Tomanek}}, \bibnamefont{and}
  \bibinfo{author}{\bibfnamefont{K.}~\bibnamefont{Richter}},
  \bibinfo{journal}{cond-mat/0709.3244}  (\bibinfo{year}{2007}).

\bibitem[{\citenamefont{Pastawski and Medina}(2001)}]{pasta01}
\bibinfo{author}{\bibfnamefont{H.~M.} \bibnamefont{Pastawski}}
  \bibnamefont{and} \bibinfo{author}{\bibfnamefont{E.}~\bibnamefont{Medina}},
  \bibinfo{journal}{Revista Mexicana de Fizica} \textbf{\bibinfo{volume}{47
  S1}}, \bibinfo{pages}{1} (\bibinfo{year}{2001}).

\bibitem[{\citenamefont{Ji et~al.}(2003)\citenamefont{Ji, Chung, Sprinzak,
  Heiblum, Mahalu, and Shtrikman}}]{ji03}
\bibinfo{author}{\bibfnamefont{Y.}~\bibnamefont{Ji}},
  \bibinfo{author}{\bibfnamefont{Y.}~\bibnamefont{Chung}},
  \bibinfo{author}{\bibfnamefont{D.}~\bibnamefont{Sprinzak}},
  \bibinfo{author}{\bibfnamefont{M.}~\bibnamefont{Heiblum}},
  \bibinfo{author}{\bibfnamefont{D.}~\bibnamefont{Mahalu}}, \bibnamefont{and}
  \bibinfo{author}{\bibfnamefont{H.}~\bibnamefont{Shtrikman}},
  \bibinfo{journal}{Nature} \textbf{\bibinfo{volume}{422}},
  \bibinfo{pages}{415} (\bibinfo{year}{2003}).

\bibitem[{\citenamefont{Roulleau et~al.}(2007)\citenamefont{Roulleau, Portier,
  Glattli, Roche, Cavanna, Faini, Gennser, and Mailly}}]{roull07}
\bibinfo{author}{\bibfnamefont{P.}~\bibnamefont{Roulleau}},
  \bibinfo{author}{\bibfnamefont{F.}~\bibnamefont{Portier}},
  \bibinfo{author}{\bibfnamefont{D.~C.} \bibnamefont{Glattli}},
  \bibinfo{author}{\bibfnamefont{P.}~\bibnamefont{Roche}},
  \bibinfo{author}{\bibfnamefont{A.}~\bibnamefont{Cavanna}},
  \bibinfo{author}{\bibfnamefont{G.}~\bibnamefont{Faini}},
  \bibinfo{author}{\bibfnamefont{U.}~\bibnamefont{Gennser}}, \bibnamefont{and}
  \bibinfo{author}{\bibfnamefont{D.}~\bibnamefont{Mailly}},
  \bibinfo{journal}{Phys.~Rev.~B} \textbf{\bibinfo{volume}{76}},
  \bibinfo{eid}{161309} (\bibinfo{year}{2007}).

\bibitem[{\citenamefont{Novoselov et~al.}(2005)\citenamefont{Novoselov, Geim,
  Morozov, Jiang, Katsnelson, Grigorieva, Dubonos, and Firsov}}]{novos05}
\bibinfo{author}{\bibfnamefont{K.~S.} \bibnamefont{Novoselov}},
  \bibinfo{author}{\bibfnamefont{A.~K.} \bibnamefont{Geim}},
  \bibinfo{author}{\bibfnamefont{S.~V.} \bibnamefont{Morozov}},
  \bibinfo{author}{\bibfnamefont{D.}~\bibnamefont{Jiang}},
  \bibinfo{author}{\bibfnamefont{M.~I.} \bibnamefont{Katsnelson}},
  \bibinfo{author}{\bibfnamefont{I.~V.} \bibnamefont{Grigorieva}},
  \bibinfo{author}{\bibfnamefont{S.~V.} \bibnamefont{Dubonos}},
  \bibnamefont{and} \bibinfo{author}{\bibfnamefont{A.~A.}
  \bibnamefont{Firsov}}, \bibinfo{journal}{Nature}
  \textbf{\bibinfo{volume}{438}}, \bibinfo{pages}{197} (\bibinfo{year}{2005}).

\bibitem[{\citenamefont{Ando}(1991)}]{ando91}
\bibinfo{author}{\bibfnamefont{T.}~\bibnamefont{Ando}}, \bibinfo{journal}{Phys.
  Rev. B} \textbf{\bibinfo{volume}{44}}, \bibinfo{pages}{8017}
  (\bibinfo{year}{1991}).

\bibitem[{\citenamefont{Neto et~al.}(2007)\citenamefont{Neto, Guinea, Peres,
  Novoselov, and Geim}}]{neto07}
\bibinfo{author}{\bibfnamefont{A.~H.~C.} \bibnamefont{Neto}},
  \bibinfo{author}{\bibfnamefont{F.}~\bibnamefont{Guinea}},
  \bibinfo{author}{\bibfnamefont{N.~M.~R.} \bibnamefont{Peres}},
  \bibinfo{author}{\bibfnamefont{K.~S.} \bibnamefont{Novoselov}},
  \bibnamefont{and} \bibinfo{author}{\bibfnamefont{A.~K.} \bibnamefont{Geim}},
  \bibinfo{journal}{cond-mat/0709.1163}  (\bibinfo{year}{2007}).

\bibitem[{\citenamefont{Zhitenev et~al.}(2000)\citenamefont{Zhitenev, Fulton,
  Yacoby, Hess, Pfeiffer, and West.}}]{zhite05}
\bibinfo{author}{\bibfnamefont{N.~B.} \bibnamefont{Zhitenev}},
  \bibinfo{author}{\bibfnamefont{T.~A.} \bibnamefont{Fulton}},
  \bibinfo{author}{\bibfnamefont{A.}~\bibnamefont{Yacoby}},
  \bibinfo{author}{\bibfnamefont{H.~F.} \bibnamefont{Hess}},
  \bibinfo{author}{\bibfnamefont{L.~N.} \bibnamefont{Pfeiffer}},
  \bibnamefont{and} \bibinfo{author}{\bibfnamefont{K.}~\bibnamefont{West.}},
  \bibinfo{journal}{Nature} \textbf{\bibinfo{volume}{404}},
  \bibinfo{pages}{473} (\bibinfo{year}{2000}).

\end{thebibliography}

\end{document}